\documentclass[aps,pra,preprint,superscriptaddress,showpacs]{revtex4-1}

\usepackage{amsmath,amssymb,amsfonts}
\usepackage[english]{babel}
\usepackage{graphicx}
\usepackage{bbm}

\begin{document}

%
\title{Ultracold two-level atom in a quadratic potential}

\author{F. Soto-Eguibar}
\affiliation{Instituto Nacional de Astrof\'{\i}sica, \'Optica y Electr\'onica, Calle Luis Enrique Erro No. 1, Sta. Ma. Tonantzintla, Pue. CP 72840, M\'exico.}

\author{A. Z\'u\~niga-Segundo}
\affiliation{Instituto Nacional de Astrof\'{\i}sica, \'Optica y Electr\'onica, Calle Luis Enrique Erro No. 1, Sta. Ma. Tonantzintla, Pue. CP 72840, M\'exico.}

\author{B.~M. Rodr\'{\i}guez-Lara}
\affiliation{Instituto Nacional de Astrof\'{\i}sica, \'Optica y Electr\'onica, Calle Luis Enrique Erro No. 1, Sta. Ma. Tonantzintla, Pue. CP 72840, M\'exico.}
\email{bmlara@inaoep.mx}

\author{H.~M. Moya-Cessa}
\affiliation{Instituto Nacional de Astrof\'{\i}sica, \'Optica y Electr\'onica, Calle Luis Enrique Erro No. 1, Sta. Ma. Tonantzintla, Pue. CP 72840, M\'exico.}

%
\begin{abstract}
We use a right unitary decomposition to study an ultracold two-level atom interacting with a quantum field.
We show that such a right unitary approach simplifies the numerical evolution for arbitrary position-dependent atom-field couplings. 
In particular, we provide a closed form, analytic time evolution operator for atom-field couplings with quadratic dependence on the position of the atom. 
\end{abstract}


\maketitle

%
\section{Introduction}

The Jaynes-Cummings (JC) model describing the interaction of a two-level atom with a quantized field mode \cite{Jaynes1963p89} is a solvable working model of the micromaser \cite{Meschede1985p551}. 
In this model, the center of mass velocity of the two-level atom is slow enough to allow controlled atom by atom interaction with the field but fast enough to be described by classical physics; e.g. thermal Rydberg atoms passing through a superconducting cavity showing Rabi oscillations \cite{Brune1996p1800}. 
In the limit case of a two-level atom so slow that its center of mass motion needs to be quantized, the system is described by the following Hamiltonian \cite{Englert1991p25}, 
\begin{eqnarray}
\hat{H} = \frac{1}{2} \hat{p}^{2} +  \omega \hat{a}^{\dagger} \hat{a} + \frac{\omega_{q}}{2} \hat{\sigma}_{z} + g(\hat{z}) \left( \hat{a}^{\dagger} \hat{\sigma}_{-} + \hat{a} \hat{\sigma}_{+} \right),
\end{eqnarray} 
where the quantized motion of the two-level atom with unitary mass has been taken in the $\hat{z}$-direction with associated canonical momentum $\hat{p}$, the quantum field is described by the annihilation (creation) operators $\hat{a}$ ($\hat{a}^{\dagger}$) and the frequency $\omega$, and the inner state of the two-level atom by the Pauli matrices $\hat{\sigma}_{j}$ with $j=z,+,-$ and the transition frequency $\omega_{q}$.
Two regimes of interest can be identified for this model, depending on the ratio between the atomic kinetic energy and the field-atom interaction energy \cite{Meyer1997p4142}: the intermediate regime, where the mean atomic kinetic energy is of the order of the mean field-atom interaction energy, and the mazer regime, where the kinetic energy is smaller.
Amplification via $z$-motion induced emission of radiation occurs in the latter and gives origin to the mazer name \cite{Scully1996p4144,Meyer1997p4142,Loffler1997p4153,Schroder1997p4164}.
This model is of interest as cavity quantum electrodynamics (cavity-QED) experiments in these two regimes appear feasible with microwave and optical quantum fields \cite{Loffler1997p4153,Walther2006p1325,Hood2000p1447}.
Also, it is feasible to control or switch off spin interactions of ultracold atoms trapped in optical lattices \cite{Duan2003p090402}, as well as to address individual sites of such lattices \cite{Bakr2009p74,Weitenberg2011p319} at the moment and, in the near future, it may be possible to couple an individual site to a quantum field as cavity-QED has been demonstrated with Bose-Einstein condensates \cite{Brennecke2007p268,Colombe2007p272}.

In the theoretical side of the problem, analytic solutions are known for the mesa, $\mathrm{sech}^{2}$, and sinusoidal modes \cite{Meyer1997p4142,Loffler1997p4153}. 
Also, an adiabatic approximation has been proposed for sinusoidal and Gaussian modes \cite{Larson2006p033805}.
Here, we introduce a right unitary approach to the problem and provide an analytic solution for a quadratic mode.
A quadratic mode may be related to an ultracold two-level atom approaching the maximum of a cavity field in an oblique path or trapped in a sinusoidal optical lattice.
In the following section, we introduce the right unitary decomposition of the model Hamiltonian for a general quantum field and construct its time evolution operator. 
Then, we study the resonant case for quadratic couplings and provide a closed form analytic time evolution operator for the system.
Finally, we study the interaction of an ultracold excited atom with number and coherent states of the quantum field.

\section{Right unitary decomposition}

By moving into the frame defined by the excitation number, $\hat{a}^{\dagger} \hat{a} + \hat{\sigma}_{z} /2 $, rotating at frequency $\omega$, we obtain an interaction picture Hamiltonian,
\begin{eqnarray}
\hat{H}_{I} &=& \frac{1}{2} \hat{p}^2  + \frac{\delta}{2} \hat{\sigma}_{z} + g(\hat{z}) \left( \hat{a}^{\dagger} \hat{\sigma}_{-} + \hat{a} \hat{\sigma}_{+} \right),
\end{eqnarray} 
where the parameter $\delta = \omega_{q} - \omega$ is the detuning between the two-level atom and field frequencies.
We can follow a right unitary approach \cite{RodriguezLara2011p3770,RodriguezLara2013p095301} to decompose this Hamiltonian into the following product, 
\begin{eqnarray}
\hat{H}_{I} = \hat{T} \hat{R}_{y} \hat{H}_{z} \hat{R}_{y}^{\dagger} \hat{T}^{\dagger},
\end{eqnarray}
where we used a rotation of $\pi/4$ radians around the $\hat{\sigma}_{y}$ operator, 
\begin{eqnarray}
\hat{R}_{y} &=& e^{i \frac{\pi}{4} \hat{\sigma}_{y}}, \\
&=& \frac{1}{\sqrt{2}} \left( \begin{array}{cc} 1 & 1 \\ -1 & 1
\end{array} \right),
\end{eqnarray}
and the transformation,
\begin{eqnarray}
\hat{T} = \left( \begin{array}{cc} \hat{V} & 0 \\ 0 & 1 \end{array} \right), \quad \hat{T}^{\dagger} = \left( \begin{array}{cc} \hat{V}^{\dagger} & 0 \\ 0 & 1 \end{array} \right).
\end{eqnarray}
The latter is right unitary, $\hat{T} \hat{T}^{\dagger}= \mathbbm{1}$ and $\hat{T}^{\dagger} \hat{T} \ne \mathbbm{1}$, due to the properties of the London exponential of the phase \cite{London1927p193,Schleich2001}, also known as Susskind-Glogower \cite{Susskind1964p49}, operators, 
\begin{eqnarray}
\hat{V} = \frac{1}{\sqrt{\hat{a}\hat{a}^{\dagger}}} \hat{a}, \quad  \hat{V}^{\dagger} = \hat{a}^{\dagger} \frac{1}{\sqrt{\hat{a}\hat{a}^{\dagger}}} ,
\end{eqnarray}
that yield, in the Fock or number state basis,
\begin{eqnarray}
\hat{V} \hat{V}^{\dagger}&=& \mathbbm{1}, \\
\hat{V}^{\dagger} \hat{V}&=&  \mathbbm{1} - \vert 0 \rangle \langle 0 \vert.
\end{eqnarray}
The new auxiliary Hamiltonian is given by
\begin{eqnarray}
\hat{H}_{z} = \frac{1}{2} \hat{p}^{2} + g(\hat{z}) \sqrt{ \hat{a}^{\dagger} \hat{a} } ~\hat{\sigma}_{z} - \frac{\delta}{2} \hat{\sigma}_{x}.
\end{eqnarray}
It is straightforward to use the properties of the right unitary approach \cite{RodriguezLara2013p095301} and show that the evolution operator of the system  is given by the following expression,\begin{eqnarray}
\hat{U}_{I}(t) &=&  e^{-i \hat{H}_{I} t}, \\
&=&   \hat{T} \hat{R}_{y} e^{-i \hat{H}_{z} t} \hat{R}_{y}^{\dagger} \hat{T}^{\dagger} .
\end{eqnarray} 
In other words, the right unitary operators for this Hamiltonian behave like a unitary operator such that $\left( \hat{T} \hat{R}_{y} \hat{H}_{z} \hat{R}_{y}^{\dagger} \hat{T}^{\dagger} \right)^j =  \hat{T} \hat{R}_{y} \hat{H}_{z}^{j} \hat{R}_{y}^{\dagger} \hat{T}^{\dagger}$.
But the construction of an analytic closed form evolution operator for any given coupling function, $g(\hat{z})$, escapes our efforts.
Next, we will show that it is simple to provide such an evolution operator for quadratic potentials on-resonance.

\section{Evolution operator on-resonance for a quadratic coupling.}

For the sake of simplicity, let us consider the case of an atom and cavity field on resonance, $\delta = 0$. Note that the approach works for the off-resonant case too, and for a quadratic potential,
\begin{eqnarray}
g_{\pm}(\hat{z}) = g_{0} \pm \frac{\vert \lambda \vert}{2} \hat{z}^2, 
\end{eqnarray}
we can write the auxiliary Hamiltonian in the following form, 
\begin{eqnarray}
\hat{H}_{z,\pm} &=& \left( \begin{array}{cc} \frac{1}{2} \hat{p}^{2} +  \sqrt{ \hat{a}^{\dagger} \hat{a} } ~ \left(g_{0} \pm \frac{\lambda}{2} \hat{z}^2 \right)& 0 \\ 0 & \frac{1}{2} \hat{p}^{2} - \frac{\lambda}{2}  \sqrt{ \hat{a}^{\dagger} \hat{a} } ~\left(g_{0} \pm \frac{\lambda}{2} \hat{z}^2 \right) \end{array} \right), \\
&=& \left( \begin{array}{cc} \hat{H}_{\pm}  +  g_{0} \sqrt{\hat{n}}& 0 \\ 0 & \hat{H}_{\mp}  - g_{0} \sqrt{\hat{n}}  \end{array} \right),
\end{eqnarray} 
where the standard, $\hat{H}_{+} = ( \hat{p}^{2} + \lambda \hat{z}^{2})/2$, and inverted, $\hat{H}_{-}=( \hat{p}^{2} - \lambda \hat{z}^{2})/2$, harmonic oscillators, are equivalent to squeezed free propagation and squeezed degenerate parametric down-conversion, in that order, or equivalently,
\begin{eqnarray}
\hat{H}_{+}&=& \omega\left(\hat{n}\right) \hat{S}\left[ \frac{1}{2} \ln \omega(\hat{n}) \right] \left( \hat{b}^{\dagger} \hat{b} + \frac{1}{2} \right)   \hat{S}^{\dagger}\left[ \frac{1}{2} \ln \omega(\hat{n}) \right], \\
\hat{H}_{-}&=&-\frac{\omega\left(\hat{n}\right)}{2}  \hat{S}\left[ \frac{1}{2} \ln \omega(\hat{n}) \right]  \left( \hat{b}^{\dagger 2}  + \hat{b}^{2} \right) \hat{S}^{\dagger}\left[ \frac{1}{2} \ln \omega(\hat{n}) \right].
\end{eqnarray}
Here we defined a frequency in terms of the number operator of the field, $\hat{n}= \hat{a}^{\dagger} \hat{a}$, 
\begin{eqnarray}
\omega\left(\hat{n}\right) = \sqrt{\vert \lambda \vert \sqrt{\hat{n}}},
\end{eqnarray}
also, we used a boson representation for the atomic center of mass motion,
\begin{eqnarray}
\hat{b} = \frac{1}{\sqrt{2}} \left( \hat{z} + i \hat{p} \right), \quad \hat{b}^{\dagger} = \frac{1}{\sqrt{2}} \left(  \hat{z} - i \hat{p}\right),
\end{eqnarray}
and the action of the squeeze operators, 
\begin{eqnarray}
\hat{S}(\hat{\xi}) = e^{- \frac{1}{2} \left( \hat{\xi} \hat{b}^{\dagger 2} - \hat{\xi}^{\dagger} \hat{b}^{2}  \right)}, 
\end{eqnarray}
where the operator $\hat{\xi}$ acts over the cavity field mode, over the position and momentum operators yields, 
\begin{eqnarray}
\hat{S}(\hat{\xi}) \hat{z} \hat{S}^{\dagger}(\hat{\xi})= \hat{z} e^{\hat{\xi}}, \quad \hat{S}(\hat{\xi}) \hat{p} \hat{S}^{\dagger}(\hat{\xi})= \hat{p} e^{-\hat{\xi}}.
\end{eqnarray}
Note that each and every Fock state of the field, $\vert k \rangle_{f}$, defines a bipartite center of mass-field mode, $\left\{ \vert j \rangle_{CM} \vert k \rangle_{f} \right\}$ with $j=0,1,2,\ldots$ and auxiliary frequency $\omega(k)=\sqrt{\vert \lambda \vert \sqrt{k}}$.
As stated in the previous section, we can easily construct the evolution operator, 
\begin{eqnarray}
\hat{U}_{I}(t) &=& \hat{T} \hat{R}_{y}  \left( \begin{array}{cc} e^{- i g_{0} \sqrt{\hat{n}} t}  \hat{S}_{\pm}(\hat{n},t)  & 0 \\ 0 & e^{ i g_{0} \sqrt{\hat{n}} t} \hat{S}_{\mp}(\hat{n},t) \end{array} \right)  \hat{R}_{y}^{\dagger} \hat{T}^{\dagger}, 
\end{eqnarray}
with the auxiliary operators, 
\begin{eqnarray}
\hat{S}_{+}(\hat{n},t) &=&  \hat{S}\left[ \frac{1}{2} \ln \omega(\hat{n}) \right] e^{- i \omega\left( \hat{n} \right) \left[ \hat{b}^{\dagger} \hat{b} + \frac{1}{2}\right]  t} \hat{S}^{\dagger}\left[ \frac{1}{2} \ln \omega(\hat{n}) \right] , \\ 
\hat{S}_{-}(\hat{n},t) &=& \hat{S}\left[ \frac{1}{2} \ln \omega(\hat{n}) \right]  e^{- \frac{i}{2} \omega\left( \hat{n} \right) \left[ \hat{b}^{\dagger 2} + \hat{b}^{2} \right]  t} \hat{S}^{\dagger}\left[ \frac{1}{2} \ln \omega(\hat{n}) \right].
\end{eqnarray}
It is possible to use the fact that the operator triad $\left\{ \hat{b}^{\dagger} \hat{b}/2 + 1/4 ,\hat{b}^{\dagger 2} /2, \hat{b}/2  \right\}$ form the $SU(1,1)$ group, in order to recast these operators as generalized squeezing operators \cite{Ban1993p1347}, 
\begin{eqnarray}
\hat{S}_{\pm}(\hat{n},t) &=&  e^{\hat{f}_{\pm}(\hat{n},t) b^{\dagger 2}} e^{\ln \hat{h}_{\pm}(\hat{n},t) \left( 2 \hat{b}^{\dagger} \hat{b} + 1\right)}   e^{\hat{f}_{\pm}(\hat{n},t) b^{2}},
\end{eqnarray}
where the sets of auxiliary operator functions are,
\begin{eqnarray}
\hat{f}_{+}(\hat{n},t) &=& \frac{ 1 - \omega^{2} \left( \hat{n} \right)  }{ 2 \left\{ 1 + \omega^{2} \left( \hat{n} \right)  - 2 i \omega\left( \hat{n} \right)  \cot \left[ \omega\left( \hat{n} \right)  t \right] \right\}}, \\
\hat{h}_{+}(\hat{n},t) &=& \sqrt{ \frac{ 2 \omega\left( \hat{n} \right) }{ 2 \omega\left( \hat{n} \right)  \cos \left[ \omega\left( \hat{n} \right) t \right] + i \left[ 1 +\omega^{2} \left( \hat{n} \right) \right] \sin \left[ \omega\left( \hat{n} \right) t \right] } },
\end{eqnarray}
and 
\begin{eqnarray}
\hat{f}_{-}(\hat{n},t) &=& \frac{ 1 + \omega^{2} \left( \hat{n} \right)  }{ 2 \left\{ 1 - \omega^{2} \left( \hat{n} \right)  - 2 i \omega\left( \hat{n} \right)  \coth \left[ \omega\left( \hat{n} \right)  t \right] \right\}}, \\
\hat{h}_{-}(\hat{n},t) &=& \sqrt{ \frac{ 2 \omega\left( \hat{n} \right) }{ 2 \omega\left( \hat{n} \right)  \cosh \left[ \omega\left( \hat{n} \right) t \right] + i \left[ 1 -\omega^{2} \left( \hat{n} \right) \right] \sinh \left[ \omega\left( \hat{n} \right) t \right] } }.
\end{eqnarray}
Again, we want to emphasize the fact that each and every number state defines a bipartite mode involving the atomic center of mass motion and that particular field state.
Thus, these operators provide a phase and squeezing for each of these modes.

\section{An excited atom interacting with number and coherent states}

Let us assume an ultra-slow atom that crosses a cavity near the maximum of the trapped coherent field in a linear $z$-trajectory. 
This allows us to approximate the field-two-level atom coupling by a quadratic function on $z$.
In this case, it is possible to describe the coupling as a quadratic potential and the evolution of the system is given by 
\begin{eqnarray}
\vert \psi(t) \rangle &=& \hat{U}_{I}(t) \vert \psi(0) \rangle.
\end{eqnarray}
In the most general case, we can consider a two-level atom in a superposition of excited and ground states entering the cavity at $z_{0}$, with center of mass linear momentum $p_{0}$, and consider some general field state in the cavity, 
\begin{eqnarray}
\vert \psi(0) \rangle = \left( \begin{array}{c} c_{e} \\ c_{g}\end{array} \right) \left\vert \beta \right\rangle_{CM} \vert \phi \rangle_{f},
\end{eqnarray}  
with $\beta = \frac{1}{\sqrt{2}}\left( z_{0} + i p_{0} \right)$, $\vert c_{e} \vert^{2} + \vert c_{g} \vert^{2} =1 $, and the center of mass is a coherent state,
\begin{eqnarray}
\vert \zeta \rangle = e^{- \frac{\vert \zeta \vert^{2}}{2}} \sum_{n} \frac{\zeta^{n}}{\sqrt{n!}} \vert n \rangle.
\end{eqnarray}

A practical example is to consider the two-level atom in the excited state, $c_{e}=1$ and $c_{g}=0$; then, it is straightforward to calculate quantities of interest, such as the mean value of the two-level atomic inversion,
\begin{eqnarray}
\langle \hat{\sigma}_{z}(t) \rangle &=& \mathrm{Re} \left[_{f}\langle \phi \vert _{CM}\langle\beta \vert e^{ 2 i g_{0} \sqrt{\hat{n}+1} t}  \hat{S}_{\pm}^{\dagger} \left( \hat{n}+1, t \right) \hat{S}_{\mp} \left( \hat{n}+1, t \right) \vert \beta \rangle_{CM} \vert \phi \rangle_{f} \right].
\end{eqnarray} 
It is also possible to calculate analytic expressions for the mean position, $\langle \hat{z}(t) \rangle$, momentum, $\langle \hat{p}(t) \rangle$, or even the $Q$-function of the field but they are not as compact as that of the mean atomic inversion.
The simplest case for this scenario is given by a field in a Fock state, 
\begin{eqnarray}
\vert \psi(0) \rangle = \vert e \rangle \vert \beta \rangle_{CM} \vert n \rangle_{f},
\end{eqnarray}
with atomic population inversions, 
\begin{eqnarray}
\langle \hat{\sigma}_{z}(t) \rangle &=& \mathrm{Re} \left[ e^{ 2 i g_{0} \sqrt{n+1} t} e^{- \vert \beta \vert^{2}} \sum_{j,k=0}^{\infty} \frac{\beta^{\ast j} \beta^{k}}{\sqrt{j!k!}}  e^{i \omega(n+1)(j+1/2)t}  \times \right. \nonumber \\ 
&& \left. \times   _{f}\langle n \vert _{CM}\langle j \vert \hat{S}_{-} \left( \hat{n}+1, t \right) \vert k \rangle_{CM} \vert n  \rangle_{f} \right], \nonumber \\
\end{eqnarray}
for the potential $g_{+}(\hat{z})$, and 
\begin{eqnarray}
\langle \hat{\sigma}_{z}(t) \rangle &=& \mathrm{Re} \left[ e^{ 2 i g_{0} \sqrt{n+1} t} e^{- \vert \beta \vert^{2}} \sum_{j,k=0}^{\infty} \frac{\beta^{\ast j} \beta^{k}}{\sqrt{j!k!}}  e^{i \omega(n+1)(k+1/2)t} \times  \right. \nonumber \\
&& \times \left.  _{f}\langle n \vert _{CM}\langle j \vert \hat{S}_{-} \left( \hat{n}+1, t \right) \vert k \rangle_{CM} \vert n \rangle_{f}  \right], \nonumber \\ 
\end{eqnarray}
for the potential $g_{-}(\hat{z})$.
Note that the term $_{f}\langle n \vert _{CM}\langle j \vert \hat{S}_{-} \left( \hat{n}+1, t \right) \vert k \rangle_{CM} \vert n \rangle_{f}$ can be calculated exactly and even approximated for large photon numbers by following \cite{deOliveira90}. 

\begin{figure}
\center\includegraphics[scale=1]{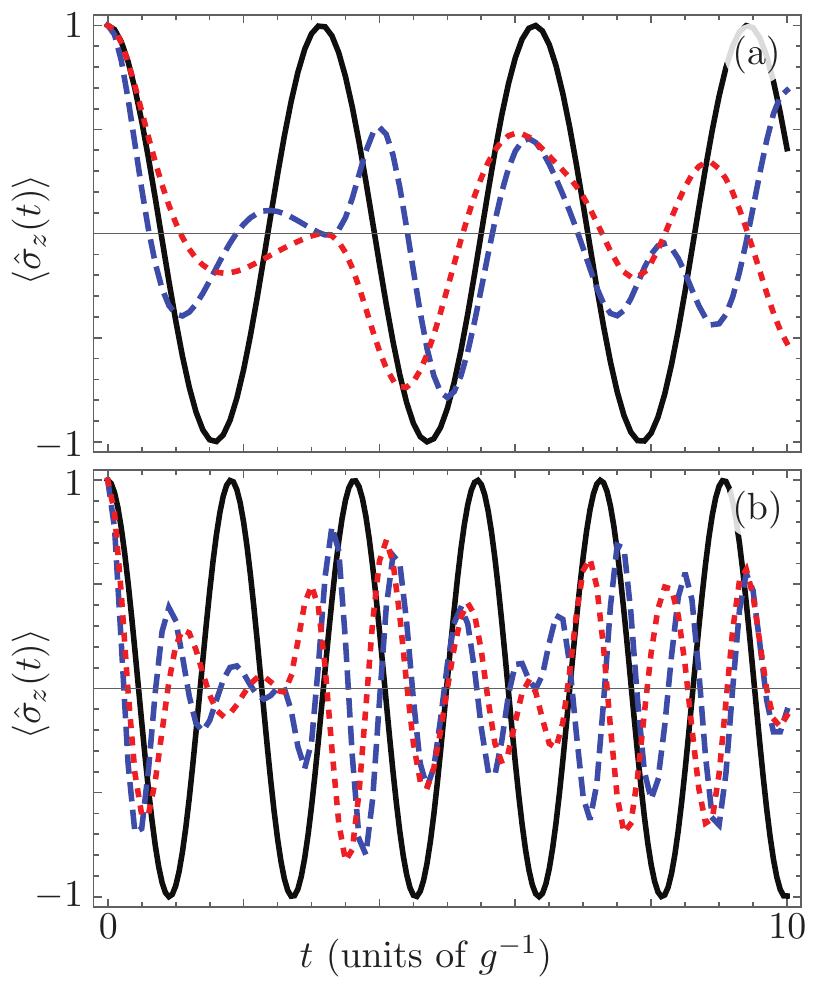}
\caption{(Color online) Time evolution of the population inversion for initial states (a) $\vert \psi(0) \rangle = \vert e \rangle \vert \beta \rangle_{CM} \vert 0 \rangle_{f}$ with $\beta = \left( -0.25 + i ~ 0.25 \right) / \sqrt{2}$ and (b) $\vert \psi(0) \rangle = \vert e \rangle \vert \beta \rangle_{CM} \vert 2 \rangle_{f}$ with $\beta = \left( -0.25 + i ~ 0.15 \right) / \sqrt{2}$ under the potentials $g_{+}(\hat{z})$ (dashed blue) and $g_{-}(\hat{z})$ (dotted red)  with paramenters $g_{0} = 1~g$ and $\lambda = 1~g$ with $g/(2 \pi) = 16 MHz$. The Rabi oscillations given by the evolution under Jaynes-Cummings dynamics is also presented (solid black).}
\label{fig:Fig1}
\end{figure}

\begin{figure}
\center\includegraphics[scale=1]{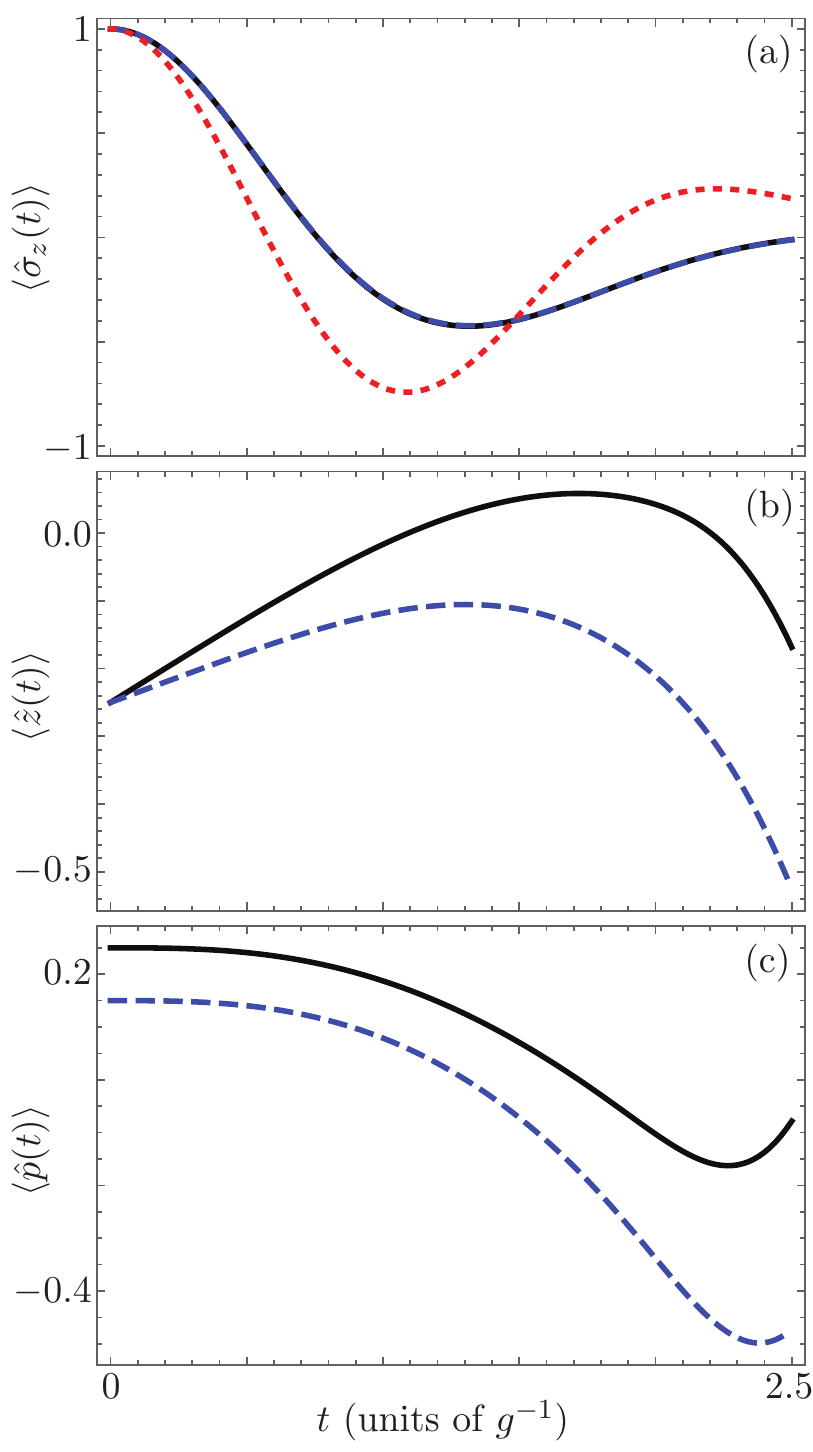}
\caption{(Color online) Time evolution of the mean value of (a) population inversion, (b) position (units of $\sqrt{\hbar / (m g)}$) and (c) momentum (units of $\sqrt{\hbar m g}$) of the atomic center of mass under the potential $g_{-}(\hat{z})$ with paramenters $g_{0} = 1~g$ and $\lambda = 1~g$. The initial states are $\vert \psi(0) \rangle = \vert e \rangle \vert \beta \rangle_{CM} \vert \alpha \rangle_{f}$ with coherent parameters $\alpha = 1$ and $\beta = \left( -0.25 + i ~ p_{0} \right) / \sqrt{2}$ where $p_{0}=0.25$ (solid black) and $p_{0}=0.15$ (dashed blue). In (a) the time evolution of the population inversion under Jaynes-Cummings dynamics (dotted red) is included.}
\label{fig:Fig2}
\end{figure}

In order to produce an example related to experimental data, let us consider the information from a cavity-cooling scheme presented in \cite{Maunz2004p50} where a single $^{85}Rb$ atom is passed through a high-finesse cavity, $\mathcal{F} = 4.4 \times 10^{5}$, that provides a coupling between the cavity $\mathrm{TEM}_{00}$ mode and the $5^{2}\mathrm{S}_{1/2}F=3 \leftrightarrow 5^{2}\mathrm{P}_{3/2}F=4$ atomic transition with a strength of $g/(2 \pi) = 16 MHz$ with an interaction length of $9~\mu$m. 
For our example, we use the value of the coupling strength as our frequency unit, $g_{0}=1$, and set the square potential strength equal to that value, $\lambda = g_{0}$, for the sake of simplicity; under this assumptions a unit of scaled time is $9.9471$~ns. 
We take the field and the atomic transition frequencies on resonance, $\delta = 0$, suppose an ideal square well trap that covers the whole $z-$axis, and artificially place the atom at the initial position $z(0)=0.6819$~nm that corresponds to an initial value $z(0)=0.25$ in $\sqrt{\hbar/ (m g)}$ units. 
We assume that the atom has two initial momenta $p(0) = 0.25$ and $p(0)=0.15$ in units of $\sqrt{\hbar m g}$ that correspond to temperatures of $23.9937$ and $8.6377$~mK, in that order. 
Figure \ref{fig:Fig1} shows the evolution of the atomic population inversion for the Jaynes-Cummings model and for an ultracold atom under the potentials $g_{\pm}(\hat{z})$ with the aforementioned parameters, and the atom initially in the excited state, with a coherent center of mass state with coherent parameter $\beta = \left( -0.25 + i ~0.25 \right) / \sqrt{2}$, interacting with a vacuum cavity field, $n=0$, Fig. \ref{fig:Fig1}(a), and a slower atom interacting with a two-photon cavity field, $\beta = \left( -0.25 + i ~0.15 \right) / \sqrt{2}$ and $n=2$, Fig. \ref{fig:Fig1}(b).
Note how the quantization of the atomic center of mass motion induces changes in the dynamics, even in the presence of an empty cavity due to emission and absorption of the initial excitation in the atom.
A more realistic scenario involves the atom finding a coherent field in the cavity.
Figure \ref{fig:Fig2} shows the time evolution of the mean population inversion, position and momentum under the potential $g_{-}(z)$ and the same set of parameters above. 
Note how the differences in the population inversion are negligible between the initial condition and how the center of mass movement of the slower atom is trapped before that of the faster atom as expected.

\section{Conclusion}

We have shown that a right unitary decomposition simplifies the problem of an ultracold two-level atom interacting with a cavity field.
In general, it is feasible to use our approach to produce the exact numerical time evolution for arbitrary $z$-dependent couplings for on- and off-resonance cases. 
In particular, we show that a quadratic potential can be solved analytically both on- and off-resonance.
As an example, we provide a closed form time for the evolution operator on-resonance; here, the evolution operator allows us to calculate closed forms for the mean values of the atomic inversion. 
The time evolution of the mean intensity of the field, position and momentum of the atomic center of mass can be calculated in closed form but are complicated enough to avoid writing them here. 
We explored the evolution of an atom originally in the excited state in the presence of number and coherent states. 

We want to note that, in theory, it may possible to use our approach to deal with a generalized potential, via a power series expansion and adequate sets of transformations, but this is unfeasible in practice because the set of transformations for each and every power has to be worked out separately.

%

\begin{thebibliography}{24}
\expandafter\ifx\csname natexlab\endcsname\relax\def\natexlab#1{#1}\fi
\expandafter\ifx\csname bibnamefont\endcsname\relax
  \def\bibnamefont#1{#1}\fi
\expandafter\ifx\csname bibfnamefont\endcsname\relax
  \def\bibfnamefont#1{#1}\fi
\expandafter\ifx\csname citenamefont\endcsname\relax
  \def\citenamefont#1{#1}\fi
\expandafter\ifx\csname url\endcsname\relax
  \def\url#1{\texttt{#1}}\fi
\expandafter\ifx\csname urlprefix\endcsname\relax\def\urlprefix{URL }\fi
\providecommand{\bibinfo}[2]{#2}
\providecommand{\eprint}[2][]{\url{#2}}

\bibitem[{\citenamefont{Jaynes and Cummings}(1963)}]{Jaynes1963p89}
\bibinfo{author}{\bibfnamefont{E.~T.} \bibnamefont{Jaynes}} \bibnamefont{and}
  \bibinfo{author}{\bibfnamefont{F.~W.} \bibnamefont{Cummings}},
  \bibinfo{journal}{Proc. IEEE} \textbf{\bibinfo{volume}{51}},
  \bibinfo{pages}{89 } (\bibinfo{year}{1963}).

\bibitem[{\citenamefont{Meschede et~al.}(1985)\citenamefont{Meschede, Walther,
  and M\"uller}}]{Meschede1985p551}
\bibinfo{author}{\bibfnamefont{D.}~\bibnamefont{Meschede}},
  \bibinfo{author}{\bibfnamefont{H.}~\bibnamefont{Walther}}, \bibnamefont{and}
  \bibinfo{author}{\bibfnamefont{G.}~\bibnamefont{M\"uller}},
  \bibinfo{journal}{Phys. Rev. Lett.} \textbf{\bibinfo{volume}{54}},
  \bibinfo{pages}{551 } (\bibinfo{year}{1985}).

\bibitem[{\citenamefont{Brune et~al.}(1996)\citenamefont{Brune, Schmidt-Kaler,
  Maali, Dreyer, Hagley, Raimond, and Haroche}}]{Brune1996p1800}
\bibinfo{author}{\bibfnamefont{M.}~\bibnamefont{Brune}},
  \bibinfo{author}{\bibfnamefont{F.}~\bibnamefont{Schmidt-Kaler}},
  \bibinfo{author}{\bibfnamefont{A.}~\bibnamefont{Maali}},
  \bibinfo{author}{\bibfnamefont{J.}~\bibnamefont{Dreyer}},
  \bibinfo{author}{\bibfnamefont{E.}~\bibnamefont{Hagley}},
  \bibinfo{author}{\bibfnamefont{J.~M.} \bibnamefont{Raimond}},
  \bibnamefont{and} \bibinfo{author}{\bibfnamefont{S.}~\bibnamefont{Haroche}},
  \bibinfo{journal}{Phys. Rev. Lett.} \textbf{\bibinfo{volume}{76}},
  \bibinfo{pages}{1800 } (\bibinfo{year}{1996}).

\bibitem[{\citenamefont{Englert et~al.}(1991)\citenamefont{Englert, Schwinger,
  Barut, and Scully}}]{Englert1991p25}
\bibinfo{author}{\bibfnamefont{B.-G.} \bibnamefont{Englert}},
  \bibinfo{author}{\bibfnamefont{J.}~\bibnamefont{Schwinger}},
  \bibinfo{author}{\bibfnamefont{A.~O.} \bibnamefont{Barut}}, \bibnamefont{and}
  \bibinfo{author}{\bibfnamefont{M.~O.} \bibnamefont{Scully}},
  \bibinfo{journal}{Europhys. Lett.} \textbf{\bibinfo{volume}{14}},
  \bibinfo{pages}{25 } (\bibinfo{year}{1991}).

\bibitem[{\citenamefont{Meyer et~al.}(1997)\citenamefont{Meyer, Scully, and
  Walther}}]{Meyer1997p4142}
\bibinfo{author}{\bibfnamefont{G.~M.} \bibnamefont{Meyer}},
  \bibinfo{author}{\bibfnamefont{M.~O.} \bibnamefont{Scully}},
  \bibnamefont{and} \bibinfo{author}{\bibfnamefont{H.}~\bibnamefont{Walther}},
  \bibinfo{journal}{Phys. Rev. A} \textbf{\bibinfo{volume}{56}},
  \bibinfo{pages}{4142 } (\bibinfo{year}{1997}).

\bibitem[{\citenamefont{Scully et~al.}(1996)\citenamefont{Scully, Meyer, and
  Walther}}]{Scully1996p4144}
\bibinfo{author}{\bibfnamefont{M.~O.} \bibnamefont{Scully}},
  \bibinfo{author}{\bibfnamefont{G.~M.} \bibnamefont{Meyer}}, \bibnamefont{and}
  \bibinfo{author}{\bibfnamefont{H.}~\bibnamefont{Walther}},
  \bibinfo{journal}{Phys. Rev. Lett.} \textbf{\bibinfo{volume}{76}},
  \bibinfo{pages}{4144 } (\bibinfo{year}{1996}).

\bibitem[{\citenamefont{L\"offler et~al.}(1997)\citenamefont{L\"offler, Meyer,
  Schr\"oder, Scully, and Walther}}]{Loffler1997p4153}
\bibinfo{author}{\bibfnamefont{M.}~\bibnamefont{L\"offler}},
  \bibinfo{author}{\bibfnamefont{G.~M.} \bibnamefont{Meyer}},
  \bibinfo{author}{\bibfnamefont{M.}~\bibnamefont{Schr\"oder}},
  \bibinfo{author}{\bibfnamefont{M.~O.} \bibnamefont{Scully}},
  \bibnamefont{and} \bibinfo{author}{\bibfnamefont{H.}~\bibnamefont{Walther}},
  \bibinfo{journal}{Phys. Rev. A} \textbf{\bibinfo{volume}{56}},
  \bibinfo{pages}{4153 } (\bibinfo{year}{1997}).

\bibitem[{\citenamefont{Schr\"oder et~al.}(1997)\citenamefont{Schr\"oder,
  Vogel, Schleich, Scully, and Walther}}]{Schroder1997p4164}
\bibinfo{author}{\bibfnamefont{M.}~\bibnamefont{Schr\"oder}},
  \bibinfo{author}{\bibfnamefont{K.}~\bibnamefont{Vogel}},
  \bibinfo{author}{\bibfnamefont{W.~P.} \bibnamefont{Schleich}},
  \bibinfo{author}{\bibfnamefont{M.~O.} \bibnamefont{Scully}},
  \bibnamefont{and} \bibinfo{author}{\bibfnamefont{H.}~\bibnamefont{Walther}},
  \bibinfo{journal}{Phys. Rev. A} \textbf{\bibinfo{volume}{56}},
  \bibinfo{pages}{4164 } (\bibinfo{year}{1997}).

\bibitem[{\citenamefont{Walther et~al.}(2006)\citenamefont{Walther, Varcoe,
  Englert, and Becker}}]{Walther2006p1325}
\bibinfo{author}{\bibfnamefont{H.}~\bibnamefont{Walther}},
  \bibinfo{author}{\bibfnamefont{B.~T.~H.} \bibnamefont{Varcoe}},
  \bibinfo{author}{\bibfnamefont{B.-G.} \bibnamefont{Englert}},
  \bibnamefont{and} \bibinfo{author}{\bibfnamefont{T.}~\bibnamefont{Becker}},
  \bibinfo{journal}{Rep. Prog. Phys.} \textbf{\bibinfo{volume}{69}},
  \bibinfo{pages}{1325 } (\bibinfo{year}{2006}).

\bibitem[{\citenamefont{Hood et~al.}(2000)\citenamefont{Hood, Lynn, Doherty,
  Parkins, and Kimble}}]{Hood2000p1447}
\bibinfo{author}{\bibfnamefont{C.~J.} \bibnamefont{Hood}},
  \bibinfo{author}{\bibfnamefont{T.~W.} \bibnamefont{Lynn}},
  \bibinfo{author}{\bibfnamefont{A.~C.} \bibnamefont{Doherty}},
  \bibinfo{author}{\bibfnamefont{A.~S.} \bibnamefont{Parkins}},
  \bibnamefont{and} \bibinfo{author}{\bibfnamefont{H.~J.}
  \bibnamefont{Kimble}}, \bibinfo{journal}{Science}
  \textbf{\bibinfo{volume}{287}}, \bibinfo{pages}{1447 }
  (\bibinfo{year}{2000}).

\bibitem[{\citenamefont{Duan et~al.}(2003)\citenamefont{Duan, Demler, and
  Lukin}}]{Duan2003p090402}
\bibinfo{author}{\bibfnamefont{L.-M.} \bibnamefont{Duan}},
  \bibinfo{author}{\bibfnamefont{E.}~\bibnamefont{Demler}}, \bibnamefont{and}
  \bibinfo{author}{\bibfnamefont{M.~D.} \bibnamefont{Lukin}},
  \bibinfo{journal}{Phys. Rev. Lett.} \textbf{\bibinfo{volume}{91}},
  \bibinfo{pages}{090402} (\bibinfo{year}{2003}).

\bibitem[{\citenamefont{Bakr et~al.}(2009)\citenamefont{Bakr, Gillen, Peng,
  Folling, and Greiner}}]{Bakr2009p74}
\bibinfo{author}{\bibfnamefont{W.~S.} \bibnamefont{Bakr}},
  \bibinfo{author}{\bibfnamefont{J.~I.} \bibnamefont{Gillen}},
  \bibinfo{author}{\bibfnamefont{A.}~\bibnamefont{Peng}},
  \bibinfo{author}{\bibfnamefont{S.}~\bibnamefont{Folling}}, \bibnamefont{and}
  \bibinfo{author}{\bibfnamefont{M.}~\bibnamefont{Greiner}},
  \bibinfo{journal}{Nature} \textbf{\bibinfo{volume}{462}}, \bibinfo{pages}{74
  } (\bibinfo{year}{2009}).

\bibitem[{\citenamefont{Weitenberg et~al.}(2011)\citenamefont{Weitenberg,
  Endres, Sherson, Cheneau, Schau{\ss}, Fukuhara, Bloch, and
  Kuhr}}]{Weitenberg2011p319}
\bibinfo{author}{\bibfnamefont{C.}~\bibnamefont{Weitenberg}},
  \bibinfo{author}{\bibfnamefont{M.}~\bibnamefont{Endres}},
  \bibinfo{author}{\bibfnamefont{J.~F.} \bibnamefont{Sherson}},
  \bibinfo{author}{\bibfnamefont{M.}~\bibnamefont{Cheneau}},
  \bibinfo{author}{\bibfnamefont{P.}~\bibnamefont{Schau{\ss}}},
  \bibinfo{author}{\bibfnamefont{T.}~\bibnamefont{Fukuhara}},
  \bibinfo{author}{\bibfnamefont{I.}~\bibnamefont{Bloch}}, \bibnamefont{and}
  \bibinfo{author}{\bibfnamefont{S.}~\bibnamefont{Kuhr}},
  \bibinfo{journal}{Nature} \textbf{\bibinfo{volume}{471}}, \bibinfo{pages}{319
  } (\bibinfo{year}{2011}).

\bibitem[{\citenamefont{Brennecke et~al.}(2007)\citenamefont{Brennecke, Donner,
  Ritter, Bourdel, Kohl, and Esslinger}}]{Brennecke2007p268}
\bibinfo{author}{\bibfnamefont{F.}~\bibnamefont{Brennecke}},
  \bibinfo{author}{\bibfnamefont{T.}~\bibnamefont{Donner}},
  \bibinfo{author}{\bibfnamefont{S.}~\bibnamefont{Ritter}},
  \bibinfo{author}{\bibfnamefont{T.}~\bibnamefont{Bourdel}},
  \bibinfo{author}{\bibfnamefont{M.}~\bibnamefont{Kohl}}, \bibnamefont{and}
  \bibinfo{author}{\bibfnamefont{T.}~\bibnamefont{Esslinger}},
  \bibinfo{journal}{Nature} \textbf{\bibinfo{volume}{450}}, \bibinfo{pages}{268
  } (\bibinfo{year}{2007}).

\bibitem[{\citenamefont{Colombe et~al.}(2007)\citenamefont{Colombe, Steinmetz,
  Dubois, Linke, Hunger, and Reichel}}]{Colombe2007p272}
\bibinfo{author}{\bibfnamefont{Y.}~\bibnamefont{Colombe}},
  \bibinfo{author}{\bibfnamefont{T.}~\bibnamefont{Steinmetz}},
  \bibinfo{author}{\bibfnamefont{G.}~\bibnamefont{Dubois}},
  \bibinfo{author}{\bibfnamefont{F.}~\bibnamefont{Linke}},
  \bibinfo{author}{\bibfnamefont{D.}~\bibnamefont{Hunger}}, \bibnamefont{and}
  \bibinfo{author}{\bibfnamefont{J.}~\bibnamefont{Reichel}},
  \bibinfo{journal}{Nature} \textbf{\bibinfo{volume}{450}}, \bibinfo{pages}{272
  } (\bibinfo{year}{2007}).

\bibitem[{\citenamefont{Larson and Stenholm}(2006)}]{Larson2006p033805}
\bibinfo{author}{\bibfnamefont{J.}~\bibnamefont{Larson}} \bibnamefont{and}
  \bibinfo{author}{\bibfnamefont{S.}~\bibnamefont{Stenholm}},
  \bibinfo{journal}{Phys. Rev. A} \textbf{\bibinfo{volume}{73}},
  \bibinfo{pages}{033805} (\bibinfo{year}{2006}).

\bibitem[{\citenamefont{Rodr{\'\i}guez-Lara
  et~al.}(2011)\citenamefont{Rodr{\'\i}guez-Lara, Rodr{\'i}guez-M{\'e}ndez, and
  Moya-Cessa}}]{RodriguezLara2011p3770}
\bibinfo{author}{\bibfnamefont{B.~M.} \bibnamefont{Rodr{\'\i}guez-Lara}},
  \bibinfo{author}{\bibfnamefont{D.}~\bibnamefont{Rodr{\'i}guez-M{\'e}ndez}},
  \bibnamefont{and}
  \bibinfo{author}{\bibfnamefont{H.}~\bibnamefont{Moya-Cessa}},
  \bibinfo{journal}{Phys. Lett. A} \textbf{\bibinfo{volume}{375}},
  \bibinfo{pages}{3770 } (\bibinfo{year}{2011}).

\bibitem[{\citenamefont{Rodr{\'\i}guez-Lara and
  Moya-Cessa}(2013)}]{RodriguezLara2013p095301}
\bibinfo{author}{\bibfnamefont{B.~M.} \bibnamefont{Rodr{\'\i}guez-Lara}}
  \bibnamefont{and} \bibinfo{author}{\bibfnamefont{H.~M.}
  \bibnamefont{Moya-Cessa}}, \bibinfo{journal}{J. Phys. A: Math. Theor.}
  \textbf{\bibinfo{volume}{46}}, \bibinfo{pages}{095301}
  (\bibinfo{year}{2013}).

\bibitem[{\citenamefont{London}(1927)}]{London1927p193}
\bibinfo{author}{\bibfnamefont{F.}~\bibnamefont{London}}, \bibinfo{journal}{Z.
  Phys.} \textbf{\bibinfo{volume}{40}}, \bibinfo{pages}{193 }
  (\bibinfo{year}{1927}).

\bibitem[{\citenamefont{Schleich}(2001)}]{Schleich2001}
\bibinfo{author}{\bibfnamefont{W.~P.} \bibnamefont{Schleich}},
  \emph{\bibinfo{title}{Quantum optics in phase state}}
  (\bibinfo{publisher}{Wiley-VCH}, \bibinfo{address}{Berlin},
  \bibinfo{year}{2001}).

\bibitem[{\citenamefont{Susskind and Glogower}(1964)}]{Susskind1964p49}
\bibinfo{author}{\bibfnamefont{L.}~\bibnamefont{Susskind}} \bibnamefont{and}
  \bibinfo{author}{\bibfnamefont{J.}~\bibnamefont{Glogower}},
  \bibinfo{journal}{Physics} \textbf{\bibinfo{volume}{1}}, \bibinfo{pages}{49 }
  (\bibinfo{year}{1964}).

\bibitem[{\citenamefont{Ban}(1993)}]{Ban1993p1347}
\bibinfo{author}{\bibfnamefont{M.}~\bibnamefont{Ban}}, \bibinfo{journal}{J.
  Opt. Soc. Am. B} \textbf{\bibinfo{volume}{10}}, \bibinfo{pages}{1347 }
  (\bibinfo{year}{1993}).

\bibitem[{\citenamefont{de~Oliveira et~al.}(1990)\citenamefont{de~Oliveira,
  Kim, Knight, and Bu\v{z}ek}}]{deOliveira90}
\bibinfo{author}{\bibfnamefont{F.~A.~M.} \bibnamefont{de~Oliveira}},
  \bibinfo{author}{\bibfnamefont{M.~S.} \bibnamefont{Kim}},
  \bibinfo{author}{\bibfnamefont{P.~L.} \bibnamefont{Knight}},
  \bibnamefont{and}
  \bibinfo{author}{\bibfnamefont{V.}~\bibnamefont{Bu\v{z}ek}},
  \bibinfo{journal}{Phys. Rev. A} \textbf{\bibinfo{volume}{41}},
  \bibinfo{pages}{2645 } (\bibinfo{year}{1990}).

\bibitem[{\citenamefont{Maunz et~al.}(2004)\citenamefont{Maunz, Puppe,
  Schuster, Syassen, Pinkse, and Rempe}}]{Maunz2004p50}
\bibinfo{author}{\bibfnamefont{P.}~\bibnamefont{Maunz}},
  \bibinfo{author}{\bibfnamefont{T.}~\bibnamefont{Puppe}},
  \bibinfo{author}{\bibfnamefont{I.}~\bibnamefont{Schuster}},
  \bibinfo{author}{\bibfnamefont{N.}~\bibnamefont{Syassen}},
  \bibinfo{author}{\bibfnamefont{P.~W.~H.} \bibnamefont{Pinkse}},
  \bibnamefont{and} \bibinfo{author}{\bibfnamefont{G.}~\bibnamefont{Rempe}},
  \bibinfo{journal}{Nature} \textbf{\bibinfo{volume}{428}}, \bibinfo{pages}{50
  } (\bibinfo{year}{2004}).

\end{thebibliography}

\end{document}